\documentclass[12pt]{iopart}

\usepackage{iopams}
\usepackage{epsfig}
\usepackage{graphicx, subfigure}
\usepackage{multirow}
\usepackage{hyperref}
\usepackage{cite}
\usepackage{color}
\def\nn{\nonumber}

\begin{document}

\title[2D npDVR on np grids]{2D Nondirect Product Discrete Variable Representation for Schr\"odinger Equation with Nonseparable Angular Variables}

\author{Sara Shadmehri$^{1}$ , Shahpoor Saeidian$^{2}$, and Vladimir S. Melezhik$^{1,3}$}
\address{$^1$ Bogoliubov Laboratory of Theoretical Physics, Joint Institute for Nuclear Research, Dubna, Moscow Region 141980, Russian Federation}
\address{$^2$ Optics and Photonics Research Center, Department of Physics, Institute for Advanced Studies in Basic Sciences (IASBS), Gava Zang, Zanjan 45137-66731, Iran}
\address{$^3$ Dubna State University, 19 Universitetskaya St., Moscow Region 141982, Russian Federation}

\eads{\mailto{shadmehri@theor.jinr.ru}, \mailto{saeidian@iasbs.ac.ir}, \mailto{melezhik@theor.jinr.ru}}

\vspace{10pt}
\begin{indented}
\item[]August 2019
\end{indented}



\begin{abstract}
We develop a nondirect product discrete variable representation (npDVR) for treating quantum dynamical problems which involve nonseparable angular variables. The npDVR basis is constructed on spherical functions orthogonalized on the grids of the Lebedev or Popov 2D quadratures for the unit sphere instead of the direct product of 1D quadrature rules. We compare our computational scheme with the old one that used the product of 1D Gaussian quadratures in terms of their convergence and  efficiency by calculating, as an example, the spectrum of a hydrogen atom in the magnetic and electric fields arbitrarily oriented to one another.  The use of the npDVR based on the Lebedev or Popov 2D quadratures substantially accelerates the convergence of the computational scheme. Moreover, we get the fastest convergence with the npDVR based on the Popov quadratures, which has the largest efficiency coefficient.
\end{abstract}


\section{Introduction}
The problem of integration of the Schr\"odinger equation with the nonseparable angular part has numerous applications. The conventional partial wave expansion becomes inefficient here due to the strong coupling of the partial waves, and therefore special computational techniques are required. In a series of papers \cite{Melezhik91,Melezhik93,Melezhik97,Melezhik98,Melezhik99} an alternative approach was suggested and developed where the angular part of the Schr\"odinger equation was approximated with the 2D nondirect product discrete variable representation (npDVR)\footnote{The 2D DVR angular basis suggested in \cite{Melezhik97,Melezhik98,Melezhik99} was later called the nondirect product DVR (npDVR) in the work of Wang and Carrington \cite{Wang}}. This approach proved to be very efficient with respect to this class of tasks, and a number of topical problems of quantum physics were resolved \cite{Melezhik93,Melezhik99,Melezhik2000,Melezhik2001,Capel,Melezhik2003,Kim,Saeidian,Schmelcher,Melezhik2011,Giannakeas,Shadmehri,Melezhik2014}.
The npDVR turned out to be very efficient in solving the time-dependent Schr\"odinger equation with three \cite{Melezhik97,Melezhik98,Melezhik99,Melezhik2001,Capel,Melezhik2003} and four \cite{Kim,Schmelcher,Melezhik2007} spatial variables.
In the present work, we further develop the computational scheme by using the Lebedev \cite{Lebedev75, Lebedev76, Lebedev77, Lebedev92, Lebedev99} and Popov \cite{Popov94} 2D quadratures on the unit sphere for constructing the npDVR basis instead of the direct product of 1D Gauss quadratures used so far.

We should note that to get an efficient DVR for the nonseparable angular part of the Schr\"odinger equation still remains a relevant problem \cite{Yu2017},  despite the intensive work to develop multi-dimensional DVRs of different kinds for molecular dynamics with direct and nondirect product (generally pruned) basis, based on Gaussian, Lebedev or Smolyak \cite{Smolyak} grids, where much progress has been achieved over the past decade \cite{Wang,Avila2009,Avila2011,Avila2011_2,Brown,Lauv2014,Larsson}. Mention also a recent work by H-G Yu \cite{Yu2017}, where coherent 2D DVR (ZDVR) method \cite{Yu2005} was extended to the unit sphere, potentially promising in order to tackle the problems in polyatomic molecular dynamics.

The discrete variable representation (DVR) \cite{Dickinson, Lill, Light} (or the Lagrange-mesh method \cite{Baye86}) proved to be useful in different quantum dynamical problems \cite{Light, Baye, Leforestier, Colbert, Corey, Sukiasyan} due to its known advantages. In DVR, any local spatial operator is diagonal, being approximated by its value at the DVR grid points. Thus, the evaluation of the Hamiltonian matrix is greatly facilitated in this representation while the only off-diagonal elements belong to the kinetic energy operator.
However, simple extension of this representation to the 2D angular subspace in the form of a direct product of two 1D DVRs is inefficient due to the singularities that arise when approximating the $\hat{L}^2$ operator in the kinetic energy \cite{Wang,Light}. As an alternative, the 2D nondirect product DVR was proposed in \cite{Melezhik91, Melezhik93, Melezhik97,Melezhik98,Melezhik99}  with the linear combinations of the spherical harmonics in the DVR basis orthogonalized on the angular 2D grids, in which the mentioned disadvantage was eliminated. However, the 2D angular grid ($\theta_j,\phi_j$) was constructed as a direct product of 1D Gaussian quadratures over the angular variables $\theta$ and $\phi$.
The discretization of the unit sphere in the 2D npDVR using the product of 1D Gaussian quadratures attracts with its simplicity. However, increasing the number of the grid points in this scheme leads to clustering of the points around the poles $x=y=0$ of the sphere, which is an essential drawback for constructing time-dependent computational schemes where the step of integration over time must be consistent with the spatial integration step. Moreover, the efficiency of a quadrature for numerical integration over the surface of the unit sphere - the ratio of the number of spherical harmonics to which the quadrature produces exact integration to the number of degrees of freedom of the quadrature \cite{McLaren} - is minimal in the 2D quadrature constructed as a direct product of 1D quadratures.

To eliminate the drawback mentioned above,
we use here the Lebedev \cite{Lebedev75, Lebedev76, Lebedev77, Lebedev92, Lebedev99} and Popov \cite{Popov94} 2D angular quadratures invariant under the octahedral and icosahedral rotation group, respectively. They belong to the class of quadrature rules (cubatures) for the unit sphere invariant under finite rotation groups outlined by Sobolev in his seminal works \cite{Sobolev,Sobolev92}.
The cubatures are constructed in the form of sets of $N$ nodal points and corresponding weights on the surface of a unit sphere, which allow precise integration of all spherical harmonics up to and including the $n$-th angular momentum, called the order of the cubature formula. Values of $N$ and $n$ determine the efficiency coefficient of every cubature formulas, $\eta={(n+1)}^2/(3N)$ \cite{McLaren}. The Lebedev and Popov cubatures pose high efficiency coefficients approaching unity with increasing $N$ while the cubature constructed as a direct product of 1D Gaussian quadratures has constant coefficient $\eta$=2/3 independently of $N$ \cite{Ahrens}. In the Lebedev and Popov cubatures, there is also no clustering of the nodal points noted above near the poles $x = y= 0$ with increasing $N$. Moreover, an outline of the implementation of the Lebedev cubatures into the 2D DVR over two angular variables was already considered by Haxton \cite{Haxton}.

In this work, we suggest a computational scheme for integration of the 3D Schr\"odinger equation with the nonseparable angular part in the 2D npDVR based on the Popov and Lebedev cubatures. We construct the orthogonalized DVR basis functions with respect to the Lebedev (or Popov) cubature grid points. Then, we compare our computational scheme with the old one that used the product of 1D Gaussian quadratures in terms of their computational efficiency by calculating, as an example, the spectrum of a hydrogen atom in crossed magnetic and electric fields, which does not permit angular variable separation. It is shown that the use of the npDVR based on the Lebedev or Popov 2D cubatures substantially accelerates the convergence of the computational scheme. Moreover, we show the superiority of the 2D npDVR based on the Popov cubatures, which has the highest efficiency coefficient $\eta$ of all the considered quadratures.



\section{3D Schr\"odinger equation with nonseparable angular variables in npDVR}
In the npDVR \cite{Melezhik2014,Melezhik2016} being used so far the solution of the Schr\"odinger equation, associated with a particle of mass $m$ under the interaction $V({\bi r})$,
\begin{eqnarray}\label{general-Sch-Eq}
\left[-\frac{\hbar^2}{2m}\nabla^2 + V({\bi r})\right]\psi({\bi r})=E\psi({\bi r})
\end{eqnarray}
is expressed as the expansion
\begin{eqnarray}\label{main-psi-expansion}
\psi(r,\theta, \phi)=\frac{1}{r}\sum_{j=1}^{N}f_j(\Omega)u_j(r)
\end{eqnarray}
in the basis
\begin{eqnarray}\label{f_j}
f_j(\Omega)=\sum_{\nu=1}^{N}\bar{Y}_\nu(\Omega)[Y^{-1}]_{\nu j}
\end{eqnarray}
 defined on the 2D angular grid $\Omega_j=(\theta_{j_\theta},\phi_{j_\phi})$ in the subspace $\Omega =(\theta, \phi)$, where $j_\theta=1,..,N_\theta~,~j_\phi=1,..,N_\phi$ and $N=N_\theta\times N_\phi$. The index $\nu$ numerates the pair $\{l,m\}$ related to the corresponding modified spherical harmonics, and the matrix $Y^{-1}$ is inverse to the matrix $Y$ defined by $Y_{j\nu}=\sqrt{w_j}\bar{Y}_{\nu}(\Omega_j)$ where $w_j=(2\pi{w_j}^\prime)/N_\phi$ shows the weight function, (${w_j}^\prime$ weight function of the Gaussian quadrature over $\theta$) and the basis functions are defined in terms of the associate Legendre polynomials $P_{l^\prime}^{m}(\theta)$ ,
\begin{eqnarray}\label{Ybar_nu}
\bar{Y}_\nu(\Omega)=\bar{Y}_{lm}(\Omega)=e^{im\phi}\sum_{l^\prime}{c_{l}^{l^\prime}}P_{l^\prime}^{m}(\theta)\,.
\end{eqnarray}
In general $c_{l}^{l^\prime}=\delta_{l l^\prime}$,  so the functions $\bar{Y}_\nu(\Omega)$ coincide with the spherical harmonics ${Y}_\nu(\Omega)$, except for a few functions of the highest values of $l$ where the functions are orthogonalized by the Gram-Schmidt method \cite{Shadmehri}. Thus, the constructed basis functions $\bar{Y}_\nu(\Omega)$ are orthogonal and complete on the grid $\Omega_j$ for each $N$.
The unknown coefficients $u_j(r)$ in expansion (\ref{main-psi-expansion}) are the values $\psi(r, \Omega_j)$ of the desired wave function $\psi(r, \Omega)$ at the grid points $\Omega_j$ multiplied by $(r\sqrt{w_j})$.

By substituting expansion (\ref{main-psi-expansion}) into (\ref{general-Sch-Eq}), a system of $N$ coupled Schr\"odinger-type equations should be solved for the unknown $N$-dimensional vector ${\bi u}(r)=\{u_j(r)\}_1^N$,
\begin{eqnarray}\label{main Schrodinger equation}
\{\hat{H}^{(0)}(r)+\hat{V}(r)\}{\bi u}(r)= E {\bi u}(r)
\end{eqnarray}
where the elements of the $N\times N$ matrices $\hat{H}^{(0)}(r)$ and $\hat{V}(r)$  are represented as
\begin{eqnarray}
H_{jj^\prime}^{(0)}(r)&=&-\frac{\hbar^2}{2m}(\delta_{jj^\prime}\frac{d^2}{dr^2}-\frac{L_{jj^{\prime}}^2}{r^2})\,,\label{H0_element}\\
V_{jj^\prime}(r)&=&V(r,\Omega_j)\delta_{jj^\prime}\,.\label{V_element_main}
\end{eqnarray}
Here, by considering $[Y^{-1}]_{\nu j}$ in (\ref{f_j}) as equal to $\sqrt{w_j}{\bar{Y}_\nu}^\ast(\Omega_{j})$, the elements of the $\hat{L}^2$ operator are defined as
\begin{eqnarray}
L_{jj^{\prime}}^2=\sum_{\nu=1}^N{\sqrt{w_jw_{j^\prime}}}\bar{Y}_\nu(\Omega_j)l(l+1){\bar{Y}_\nu}^\ast(\Omega_{j^\prime})\,.
\end{eqnarray}

\begin{table}[t]
\caption{\label{tab:Table1}The parameters of the Lebedev and Popov cubatures for each $N$, which is the number of grid points on a unit sphere. Each cubature produces exact integration of all spherical harmonics up to and including $n$-th angular momentum and all possible products of spherical harmonic pairs with angular momenta up to and including $l_{max}$.  $l_{top}$ is the highest $l$-value which is incorporated in constructing npDVR basis functions. The efficiency coefficient $\eta$ defined as $\eta={(n+1)}^2/(3N)$.}
\begin{indented}
\item[]\begin{tabular}{|c|c|c|c|c|}
\br
\multicolumn{5}{|c|}{LEBEDEV} \\
\hline
$N$ & $n$ & $l_{max}$ & $l_{top}$ & $\eta$ \\
\hline
6 &	3 &	1 &	2 & 0.89\\
14 & 5 & 2 & 4 & 0.85\\
26 & 7 & 3 & 6 & 0.82\\
38 & 9 & 4 & 6 & 0.87\\
50 & 11	& 5 & 8 & 0.96\\
86 & 15	& 7 & 10 & 0.99\\
110 & 17 & 8 & 12 & 0.98\\
\hline
\end{tabular}
\quad
\begin{tabular}{|c|c|c|c|c|}
\br
\multicolumn{5}{|c|}{POPOV} \\
\hline
$N$ & $n$ & $l_{max}$ & $l_{top}$ & $\eta$\\
\hline
6 &	3 &	1 &	2 & 0.89\\
12 & 5 & 2 & 3 & 1.00\\
22 & 7 & 3 & 5 & 0.97\\
32 & 9 & 4 & 6 & 1.04\\
48 & 11	& 5 & 7 & 1.00\\
84 & 15	& 7 & 9 & 1.02\\
104 & 17 & 8 & 11 & 1.04\\
\hline
\end{tabular}
\end{indented}
\end{table}

In what follows, we explain how the basis functions and further formulae are altered by using the Lebedev or Popov cubature angular grid points and corresponding weights instead of the Gaussian quadratures.
For any fixed $N$, the cubature of Lebedev or Popov type can exactly integrate the square of spherical harmonics up to a certain angular momentum ($l_{max}$) (Table \ref{tab:Table1}). Thus, only for the $(l_{max}+1)^2\leq N$ spherical functions with $l\leq l_{max}$ the orthogonality condition is satisfied:
\begin{eqnarray}
\fl <Y_{\nu}|Y_{\nu^\prime}>=\int{Y_{\nu}(\Omega)Y_{\nu^\prime}^{\ast}(\Omega)d\Omega}=4\pi\sum_{j=1}^{N}{w_jY_{\nu}(\Omega_j)Y_{\nu^\prime}^{\ast}(\Omega_j)}=\delta_{\nu\nu^\prime}~~,~~l\leq l_{max}
\end{eqnarray}
where $\Omega_j=(\theta_j,\phi_j)$ and $w_j$ show the grid's angular coordinates and its weights the sum of which equals unity ($\sum_{j=1}^{N}{w_j}=1.0$) (for Lebedev grids and weights see \cite{Burkardt}).
However, it follows from the above consideration (see Equations (\ref{main-psi-expansion}-\ref{Ybar_nu})) that for constructing the npDVR basis (\ref{f_j}) the number of basis functions should be equal to the number of grid points $N$. Therefore, the $N-(l_{max}+1)^2$ additional spherical harmonics with $l>l_{max}$ are required to be included into consideration and made orthogonal to others on the angular grid. Towards this aim, we use the orthogonalization procedure suggested in \cite{Haxton} instead of the Gramm-Schmidt orthogonalization.
First, the number of harmonics expands to a value $\bar{N}$ exceeding $N$ by including all the harmonics with $l$ up to and including $\bar{l}$ so that $\bar{N}=(\bar{l} +1)^2 \geq N$.
Then, we construct an overlap matrix of $\bar{N}\times \bar{N}$ dimension whose elements are defined as a scalar product of different available spherical harmonics with $l\leq \bar{l}$
\begin{eqnarray}
<Y_{\nu}|Y_{\nu^\prime}>=\int{Y_{\nu}(\Omega)Y_{\nu^\prime}^{\ast}(\Omega)d\Omega}=4\pi\sum_{i=1}^{N}{w_iY_{\nu}(\Omega_i)Y_{\nu^\prime}^{\ast}(\Omega_i)}~~~~~l,l^\prime\leq \bar{l}\,.
\end{eqnarray}
If by diagonalizing this matrix, we get exactly $N$ nonzero eigenvalues, we consider $\bar{l}$ as equal to $l_{top}$, and store the normalized corresponding eigenvectors in the matrix of $N\times \bar{N}$ dimension called $S$, where $N$ is the number of nonzero eigenvectors and $\bar{N}$ is the vector dimension. Otherwise, we increase $\bar{l}$ to $\bar{l}+1$ and subsequently repeat the above procedure until reaching the desired number $N$ of nonzero eigenvalues of the overlap matrix (10). The suitable values of $l_{top}$ obtained by this method are listed in Table \ref{tab:Table1}.

In this way we obtain a set of linear combinations of spherical harmonics
\begin{eqnarray}\label{phi_nu}
\Phi_\nu(\Omega)=\sum_{\mu=1}^{\bar{N}}S_{\nu\mu}Y_\mu(\Omega)~~,~~\nu=1,..,N\,,
\end{eqnarray}
where $\mu=\{l_\mu,m_\mu\}$ labeled in an ascending order as $\{l_\mu,m_\mu\}=\{0,0\},\{1,-1\},\{1,0\},\{1,1\},...,\{l_{top},l_{top}-1\},\{l_{top},l_{top}\} $. The set of the constructed functions $\Phi_\nu(\Omega)$ is orthogonal and complete on the cubature grid points $(\Omega_i,w_i)$ as
\begin{eqnarray}
<\Phi_\nu|\Phi_{\nu^\prime}>=4\pi\sum_{i=1}^{N}{w_i\Phi_{\nu}(\Omega_i)\Phi_{\nu^\prime}^{\ast}(\Omega_i)}=\delta_{\nu\nu^\prime}
\end{eqnarray}
and
\begin{eqnarray}
4\pi\sum_{\nu=1}^{N}{\sqrt{w_i}\sqrt{w_j}\Phi_{\nu}(\Omega_i)\Phi_{\nu}^{\ast}(\Omega_j)}=\delta_{ij}\,.
\end{eqnarray}
Thus, with $\Phi_\nu$  defined by (\ref{phi_nu}) we construct the desired 2D npDVR basis $f_j(\Omega)$ (\ref{f_j}) as
\begin{eqnarray}\label{f_j_new}
f_j(\Omega)=\sum_{\nu=1}^{N}\sqrt{4\pi w_j}\Phi_\nu(\Omega)\Phi^{*}_{\nu}(\Omega_j)\,.
\end{eqnarray}
In this representation, the $\hat{L}^2$ operator in (\ref{H0_element}) takes the following form:
\begin{eqnarray}\label{L2-operator}
L_{jj^\prime}^2=4\pi\sum_{\nu=1}^N{\sqrt{w_jw_{j^\prime}}}\sum_{\mu=1}^{\bar{N}}S_{\nu\mu}l_\mu(l_\mu+1){{Y}_\mu}(\Omega_{j}){\Phi}_\nu^\ast(\Omega_{j^\prime})\,,
\end{eqnarray}
and the interaction operator $\hat{V}(r)$ has the diagonal structure defined by Eq.(\ref{V_element_main}) in the Lebedev (or Popov) grid points $\Omega_j$.

\section{Numerical example: hydrogen atom in crossed electric and magnetic fields}

To perform a comparative analysis of the efficiency of the suggested npDVR computational schemes based on the Lebedev and Popov cubatures, we have calculated  low-lying bound states of a hydrogen atom in the electric $\bi{F}$ and magnetic $\bi{B}$ fields arbitrarily oriented to one another. It is a good example for our purpose, because in the general case of arbitrary mutual orientation of the fields the separation of angular variables is  not possible in this three-dimensional problem. Moreover, this problem has already been successfully investigated with the early version of the npDVR, which used a direct product of 1D angular Gaussian quadratures \cite{Melezhik93}.

In this problem, the interaction potential in the corresponding Schr\"odinger equation (\ref{main Schrodinger equation}) is expressed as
\begin{eqnarray}\label{Electron-potential in EM}
\hat{V}({\bi r})=-\frac{1}{r}+({\bi F}\cdot{\bi r})+\frac{1}{2}({\bi B}\cdot{\bi L})+\frac{1}{8}{[{\bi B}\times{\bi r}]}^2\,,
\end{eqnarray}
where ${\bi r}$ and ${\bi L}$ are the radius-vector and orbital angular momentum of the electron.
We consider the vectors ${\bi B}$ and ${\bi F}$, which form an arbitrary angle $\alpha$, are placed in the $xz$-plane,
\begin{eqnarray}
{\bi B}=\beta{\bi n}_z~~,~~
{\bi F}=\gamma(\sin\alpha{\bi n}_x+\cos\alpha{\bi n}_z)\,.
\end{eqnarray}
Here, the strengths $\gamma$ and $\beta$ of the electric and magnetic field are expressed in atomic units $F_0=e^5m_e^2/\hbar^4\approx5.14\times10^9$ V/cm and $B_0={(e/\hbar)}^3m_e^2c\approx2.35\times10^9$ G ($\hbar=e=m_e=1$), which we use hereafter. Thus, the interaction potential (\ref{Electron-potential in EM}) takes the form
\begin{eqnarray}\label{potential_EM}
\fl\hat{V}({\bf r})=-\frac{1}{r}+\gamma r\left({\sin\alpha\sin\theta\cos\phi+\cos\alpha\cos\theta}\right)+\frac{1}{2}({\beta L_z})+\frac{1}{8}{(\beta r\sin\theta)}^2\,.
\end{eqnarray}

In our 2D npDVR basis (\ref{f_j_new}), the $\hat{L}^2$ operator is defined as (\ref{L2-operator}) and the elements of the potential matrix (\ref{potential_EM}) are represented by the sum of the diagonal matrix (\ref{V_element_main}) plus the term related to the $\hat{L}_z$ operator,
\begin{eqnarray}\label{final-potential-matrix}
V_{jj^{\prime}}(r)&=&\left[-\frac{1}{r}+\gamma r\left({\sin\alpha\sin\theta_j\cos\phi_j+\cos\alpha\cos\theta_j}\right)+\frac{1}{8}{(\beta r\sin\theta_j)}^2\right]\delta_{jj^{\prime}}\nn\\&+&\frac{1}{2}\beta\left[4\pi\sum_{\nu=1}^N{\sqrt{w_jw_{j^\prime}}}\sum_{\mu=1}^{\bar{N}}S_{\nu\mu}m_\mu{{Y}_\mu}(\Omega_{j}){\Phi}_\nu^\ast(\Omega_{j^\prime})\right].
\end{eqnarray}
To calculate in the 2D npDVR of the bound state energy $E$ of the hydrogen atom in the crossed electric and magnetic fields, we must complement equations (\ref{main Schrodinger equation}) with zero boundary conditions at $r=0$ and $r=r_m\rightarrow \infty $
\begin{eqnarray}\label{zero-boundary}
u(0,\Omega_j)=u(r_m,\Omega_j)=0
\end{eqnarray}
and resolve the arising eigenvalue problem (\ref{main Schrodinger equation},\ref{H0_element},\ref{L2-operator},\ref{final-potential-matrix},\ref{zero-boundary}). For this purpose, we approximate the radial part of the system of equations (\ref{main Schrodinger equation}) with six-order finite differences on a quasi-uniform grid $r_i=r_m(\exp(7 x_i)-1)/(\exp(7)-1)$ (where $x_i=i/N_x$, $i=1,2,...,N_x$) suggested in \cite{Melezhik97} and use the inverse iterations in the subspace for finding the desired eigenvalues. The step of integration over $r$ and the integration boundary $r_m$ are fixed so that the introduced integration error does not exceed errors from the npDVR approximation of the angular part: $1/N_x =0.001~(0.0005) $ and $r_m=8~(20) $ for strong (weak) fields.

\begin{figure}[!]
\begin{center}
\begin{subfigure}{
\includegraphics[width=0.45\columnwidth]{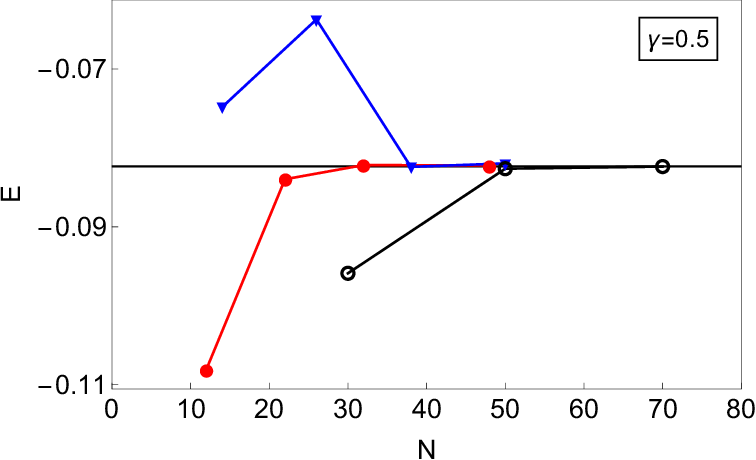} \includegraphics[width=0.45\columnwidth]{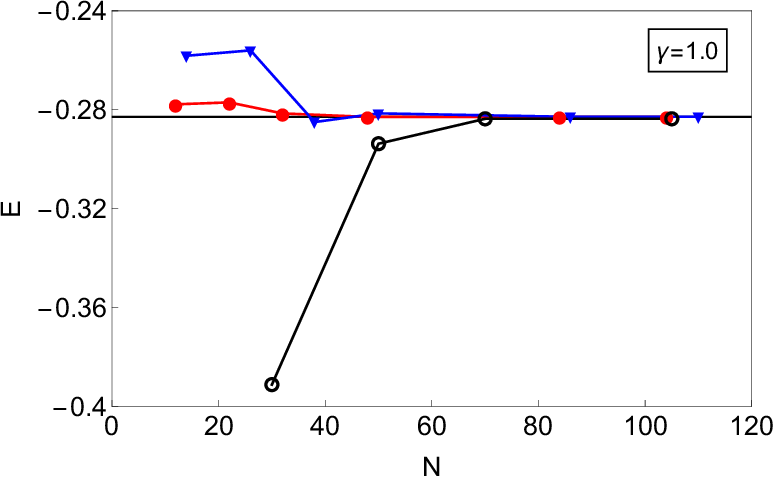}}
\end{subfigure}\\
\begin{subfigure}{
\includegraphics[width=0.45\columnwidth]{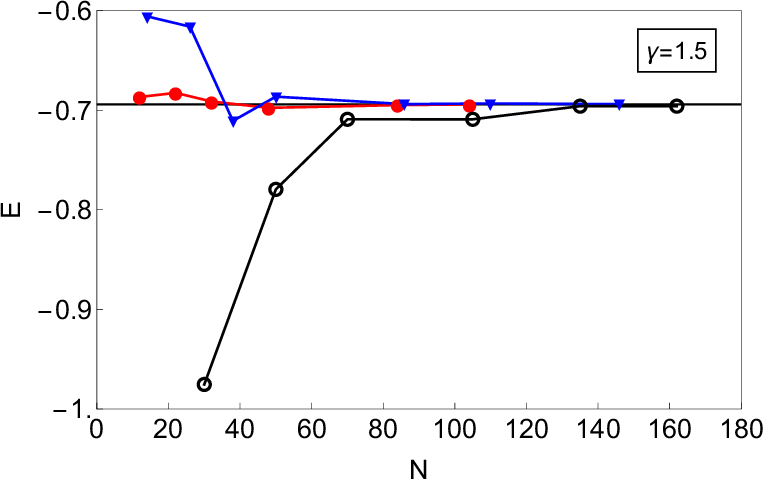} \includegraphics[width=0.45\columnwidth]{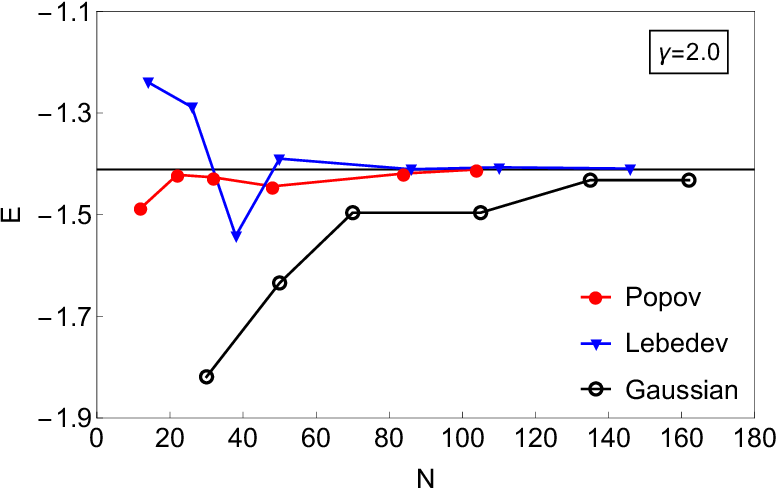}}
\end{subfigure}
\end{center}
\caption{The ground-state energies of a hydrogen atom in the external magnetic $\beta=2.0$ and  electric $\gamma=0.5, 1.0, 1.5, 2.0$ fields perpendicular to one another ($\alpha=\pi/2$). The energies are calculated using three computational schemes with different npDVR bases (see text). The thin horizontal black lines show the most accurate value obtained for each case.}
\label{fig:Fig1}
\end{figure}

First, we calculate the energies of the ground state of a hydrogen atom in the strong magnetic ($\beta=2.0$) and electric ($0.5\leq\gamma\leq 2.0$) fields at the angle $\alpha=\pi/2$. In this case, the term $\frac{1}{2}({\beta L_z})$ is absent in the potential (\ref{potential_EM}). The convergence of the calculated energies $E$ on a sequence of condensing angular grids ($N\rightarrow\infty$) for three npDVR scheme is illustrated in Figure.\ref{fig:Fig1} for a few strengths $\gamma=0.5, 1.0, 1.5, 2.0$ of the electric field. Here, the points indicated as {\it Gaussian} represent the results obtained with the npDVR based on the direct product of 1D angular Gaussian quadratures. The points indicated as {\it Lebedev} and {\it Popov} are obtained with the npDVR based on the Lebedev and Popov cubatures, respectively. The superiority of the npDVR based on the Popov and Lebedev cubatures against the old npDVR scheme that used the direct product of 1D Gaussian quadratures is obvious in the graphs. Moreover, we observe a clear correlation between the convergence of the npDVR computational scheme and the efficiency coefficient $\eta$: the larger the coefficient $\eta$, the faster the convergence of the computational scheme.

We also perform the comparative analysis of these three computational schemes as applied to the description of the splitting of the hydrogen first excited state $n=2$ in rather weak fields of magnitudes $\beta=0.005$ and $\gamma=10^{-2}/6$, which was analyzed earlier in \cite{Melezhik93}. The calculated energies of the lowest state of the multiplet with the quantum number $q=-(n-1)=-1$ \cite{Melezhik93} are presented in Table \ref{tab:table2} when the fields are perpendicular to each other ($\alpha=\pi/2$) and in Table \ref{tab:table3} when $\alpha=\pi/8$, together with the corresponding results of Melezhik \cite{Melezhik93}. It is worth mentioning that in our Gaussian scheme, we consider the basis function index $\nu=\left\{l,m\right\}$ which appears in equation (\ref{f_j}) as
\begin{eqnarray}
\sum_{\nu=1}^{N}=\sum_{m=-(N_\phi-1)/2}^{(N_\phi-1)/2}\sum_{l=\left|m\right|}^{\left|m\right|+N_\theta-1}\nn
\end{eqnarray}
while in \cite{Melezhik93} it takes the values $\nu=\{l,m\}=\{0,0\},\{1,-1\},\{1,0\},\{1,1\},...,\{\sqrt{N}-1,\sqrt{N}-2\}$ and $\{\sqrt{N}-1,\sqrt{N}-1\}$.
The comparison of the results of \cite{Melezhik93} with ours in the tables shows that our method yields accurate values and demonstrates clearly the rearrangement of the energy spectrum due to the varying angle $\alpha$. Here, similar to the case of the ground state, the performed analysis demonstrates a direct correlation between the value of the efficiency coefficient $\eta$ and the convergence of the computational scheme, i.e. applying the npDVR based on the Lebedev or Popov cubatures enhances the convergence of the computational scheme. Moreover, the fastest convergence is demonstrated by the npDVR based on the Popov cubatures, which has the largest $ \eta $ among all the considered quadratures on the unit sphere and gives rather accurate results even at a low number of basis functions as $N=12$.
\begin{table}[h]
    \caption{The energy of the lowest state $\left \{q \right \}=\left \{-1 \right \}$ of the Hydrogen atom multiplet $n=2$ in the fields $\beta=0.005$ and $\gamma=10^{-2}/6$ when $\alpha=\pi/2$. The energy is obtained using three computational schemes with different npDVR bases. In \cite{Melezhik93}, the energy was calculated to be $E=-0.13075$ by using $N=25$ grid points.}
    \label{tab:table2}
    \begin{indented}
    \item[]\begin{tabular}{|c|ccc|c|c|}
     \br
     \multicolumn{6}{|c|}{$\beta=0.005$ , $\gamma=10^{-2}/6$} \\
\hline
\multirow{2}{*}{$N$} & \multicolumn{3}{|c|}{Gaussian} & \multirow{2}{*}{Lebedev} & \multirow{2}{*}{Popov}\\
 & ${\scriptstyle N_\phi=3}$ & ${\scriptstyle N_\phi=5}$ & ${\scriptstyle N_\phi=7}$& &  \\
      \hline 
12 					& \multicolumn{3}{|c|}{---}                    & --- & -0.1307374 \\
14					& \multicolumn{3}{|c|}{---}                    & -0.1306568 & ---            \\
22 					& \multicolumn{3}{|c|}{---}                    & ---            & -0.1307374 \\
25 					&              & -0.1307454 &              & ---            & ---           \\
26 					& \multicolumn{3}{|c|}{---}                    & -0.1306159 & ---            \\
\multirow{2}{*}{30} &    -0.1326501          & 			 &    & \multirow{2}{*}{---} & \multirow{2}{*}{---}\\
					&			   & -0.1307454&               &                    &                 \\
32 					& \multicolumn{3}{|c|}{---}                    & ---            & -0.1307377 \\
36					&	-0.1326501		   &             & 		  & ---			& ---		    \\
38					& \multicolumn{3}{|c|}{---}                    & -0.1307377 & ---           \\
40					&				& -0.1307454 & 			 & ---			& ---				\\
48 					& \multicolumn{3}{|c|}{---}                    & ---            & -0.1307377 \\
49					&   &		      &		-0.1307377	     & ---		    & ---		      \\
50					&				& -0.1307454&				 & -0.1307377 & ---			\\
70					&   &			&		-0.1307377		 & ---		    & ---	         \\
      \hline 
    \end{tabular}
  \end{indented}
\end{table}
\begin{table}[h]
    \caption{The energy of the lowest state $\left \{q \right \}=\left \{-1 \right \}$ of the Hydrogen atom multiplet $n=2$ in the fields $\beta=0.005$ and $\gamma=10^{-2}/6$ when $\alpha=\pi/8$. The energy is obtained using three computational schemes with different npDVR bases. In \cite{Melezhik93}, the energy was calculated to be $E=-0.13028$ by using $N=25$ grid points. }
    \label{tab:table3}
    \begin{indented}
\item[]\begin{tabular}{|c|ccc|c|c|}
     \br
     \multicolumn{6}{|c|}{$\beta=0.005$ , $\gamma=10^{-2}/6$} \\
\hline
\multirow{2}{*}{$N$} & \multicolumn{3}{|c|}{Gaussian} & \multirow{2}{*}{Lebedev} & \multirow{2}{*}{Popov}\\
 & ${\scriptstyle N_\phi=3}$ & ${\scriptstyle N_\phi=5}$ & ${\scriptstyle N_\phi=7}$& &  \\
      \hline 
12 & \multicolumn{3}{|c|}{---} 		  & ---            & -0.1302730  \\
14 &	\multicolumn{3}{|c|}{---}		  &	-0.1301970 &	---\\
22 &	\multicolumn{3}{|c|}{---}		  &	---            &	-0.1302723\\
26 &	\multicolumn{3}{|c|}{---} 		  &	-0.1301643 &	---\\
30 & -0.1304448 &  &      &	---            &	---\\
32 &	\multicolumn{3}{|c|}{---}         &	---            &	-0.1302728\\
36 & -0.1304448 &  &      &	---            &	---\\
38 &	\multicolumn{3}{|c|}{---}         &	-0.1302728 &	---\\
40 &    & -0.1302756 &  &	---            &	---\\
48 &	\multicolumn{3}{|c|}{---}        &	---            &	-0.1302728\\
49					&   &		      &		-0.1302755	     & ---		    & ---		      \\
50 &   & -0.1302756 &  &	-0.1302728 &	---\\
70 &   &			&		-0.1302755		 & ---		    & ---	         \\
      \hline 
    \end{tabular}
\end{indented}
\end{table}

\section{Conclusion}\label{Conclusion}
We have developed the 2D npDVR for treating quantum dynamical problems which involve nonseparable angular variables. The npDVR basis is constructed on spherical functions orthogonalized on the grids of the Lebedev or Popov cubatures for the unit sphere. We have presented a detailed description of the npDVR basis construction and comparative analysis of our computational scheme with the old one that used the direct product of 1D Gaussian quadratures, in terms of their convergence by calculating, as an example, the spectrum of a hydrogen atom in the magnetic and electric fields arbitrarily oriented to one another. The performed analysis demonstrates a direct correlation between the value of the efficiency coefficient $\eta$ and the convergence of the computational scheme. We have found that the use of the npDVR based on the Lebedev or Popov cubatures substantially accelerates the convergence of the computational scheme and have shown the superiority of the npDVR based on the Popov cubatures possessing the largest efficiency coefficient. The acceleration of the convergence of the 2D DVR with increasing the efficiency coefficient of the corresponding grid is understood. Indeed, a 2D DVR with a larger efficiency coefficient gives a better approximation to the original problem than a 2D DVR, which has less $\eta$ on the same number of basis functions (grid points). This fact should be taken into account when constructing the optimal 2D DVR for the problem under consideration: one should expect faster convergence for the DVR with larger $\eta$.

Obtaining accurate results with a lower number of the npDVR basis functions $N$ makes the developed computational scheme very promising in application to the problems involved with the Schr\"odinger equations of nonseparable angular variables. Furthermore, eliminating the angular grid points clustering near the poles $x=y=0$ with increasing $N$ in the developed npDVR, enables one to extend the computational scheme to the time-dependent Schr\"odinger equation with a few nonseparable angular variables for describing quantum dynamics of atomic systems with two active electrons.

\ack
We thank Daniel J. Haxton for valuable discussions. The work was was supported by the Russian Foundation for Basic Research, Grants No. 18-02-00673 and No. 19-02-00058 and the ``RUDN University Program 5-100''.




\section*{References}
\begin{thebibliography}{40}

\bibitem{Melezhik91} Melezhik V S 1991 {\it J. Comput. Phys.} {\bf 92} 67
\bibitem{Melezhik93} Melezhik V S 1993 {\it Phys. Rev.} A {\bf 48} 4528
\bibitem{Melezhik97} Melezhik V S 1997 {\it Phys. Lett.}  A {\bf 230} 203
\bibitem{Melezhik98} Melezhik V S 1998 {\it A Computational Method for Quantum Dynamics of Three-Dimensional Atom in Strong Fields} in {\it Atoms and Molecules in Strong External Fields} editted by Schmelcher P and Schweizer W (New York: Plenum) p~89
\bibitem{Melezhik99} Melezhik V S and Baye D 1999 {\it Phys. Rev.} C {\bf 59} 3232
\bibitem{Wang} Wang X -G and Carrington T Jr 2008 {\it J. Chem. Phys.}
{\bf 128} 194109
\bibitem{Melezhik2000} Melezhik V S and Schmelcher P 2000 {\it Phys. Rev. Lett.} {\bf 84} 1870
\bibitem{Melezhik2001} Melezhik V S and Baye D 2001 {\it Phys. Rev.} C {\bf 64} 054612
\bibitem{Capel} Capel P, Baye D and Melezhik V S 2003 {\it Phys. Rev.} C
{\bf 68} 014612
\bibitem{Melezhik2003} Melezhik V S and Hu C Yu 2003 {\it Phys. Rev. Lett.} {\bf 90} 083202
\bibitem{Kim} Kim J I, Melezhik V S and Schmelcher P 2006 {\it Phys.
Rev. Lett.} {\bf 97} 193203
\bibitem{Saeidian} Saeidian S, Melezhik V S and Schmelcher P 2008 {\it
Phys. Rev.} A {\bf 77} 042721
\bibitem{Schmelcher} Melezhik V S and Schmelcher P 2009 {\it New J. Phys.} {\bf 11} 073031
\bibitem{Melezhik2011} Melezhik V S and Schmelcher P 2011 {\it Phys. Rev.} A {\bf 84} 042712
\bibitem{Giannakeas} Giannakeas P, Melezhik V S and Schmelcher P 2011 {\it
Phys. Rev.} A {\bf 84} 023618
\bibitem{Shadmehri} Shadmehri S, Saeidian S and Melezhik V S 2016 {\it Phys. Rev.} A {\bf 93} 063616
\bibitem{Melezhik2014} Melezhik V S 2014 {\it Phys. Atom. Nucl.} {\bf 77} 446
\bibitem{Melezhik2007} Melezhik V S, Kim J I and Schmelcher P 2007 {\it Phys. Rev.} A {\bf 76} 053611
\bibitem{Lebedev75} Lebedev V I 1975 {\it Zh. Vychisl. Mat. Mat. Fiz.} {\bf 15} 48
\bibitem{Lebedev76} Lebedev V I 1976 {\it Zh. Vychisl. Mat. Mat. Fiz.} {\bf 16} 293
\bibitem{Lebedev77} Lebedev V I 1977 {\it Sib. Mat. Zh.} {\bf 18} 132
\bibitem{Lebedev92} Lebedev V I and Skorokhodov A L 1992 {\it Russ. Acad. Sci. Dokl. Math.} {\bf 45} 587
\bibitem{Lebedev99} Lebedev V I and Laikov D N 1999 {\it Russ. Acad. Sci. Dokl. Math.} {\bf 59} 477
\bibitem{Popov94} Popov A S 1994 {\it Russ. J. Numer. Anal. M.} {\bf 9} 535
\bibitem{Yu2017} Yu H-G 2017 {\it J. Chem. Phys.} {\bf 147} 094101
\bibitem{Smolyak} Smolyak S A 1963 {\it Sov. Math. Dokl.} {\bf 4} 240
\bibitem{Avila2009} Avila G and Carrington T Jr 2009 {\it J. Chem. Phys.} {\bf 131}
174103
\bibitem{Avila2011} Avila G and Carrington T Jr 2011 {\it J. Chem. Phys.} {\bf 134}
054126
\bibitem{Avila2011_2} Avila G and Carrington T Jr 2011 {\it J. Chem. Phys.} {\bf
135} 064101
\bibitem{Brown} Brown J and Carrington T Jr 2016 {\it J. Chem. Phys.} {\bf 145} 144104
\bibitem{Lauv2014} Lauvergnat D and Nauts A 2014 {\it Spectrochim. Acta A} {\bf 119} 18
\bibitem{Larsson} Larsson H R, Harke B and Tannor D J 2016 {\it J. Chem. Phys} {\bf 145} 204108
\bibitem{Yu2005} Yu H-G 2005 {\it J. Chem. Phys.} {\bf 122} 164107
\bibitem{Dickinson} Dickinson A S and Certain P R 1968 {\it J. Chem. Phys.} {\bf 49} 4209
\bibitem{Lill} Lill J V, Parker G A and Light J C 1982 {\it Chem. Phys.
Lett.} {\bf 89} 483
\bibitem{Light} Light J C and Carrington T Jr 2000 {\it Adv. Chem. Phys.}
{\bf 114} 263
\bibitem{Baye86} Baye D and Heenen P-H 1986 {\it J. Phys.} A {\bf 19} 2041
\bibitem{Baye} Baye D 2006 {\it Phys. Status Solidi} B {\bf 243} 1095
\bibitem{Leforestier} Leforestier C  1991 {\it J. Chem. Phys.} {\bf 94} 6388
\bibitem{Colbert} Colbert D T and Miller W H 1991 {\it J. Chem. Phys.} {\bf 96} 1982
\bibitem{Corey} Corey G C and Tromp J W 1995 {\it J. Chem. Phys.} {\bf 103} 1812
\bibitem{Sukiasyan} Sukiasyan S and Meyer H-D 2001 {\it J. Phys. Chem.} A {\bf 105} 2604
\bibitem{McLaren} McLaren A D 1963 {\it Math. Comput.} {\bf 17}, 361
\bibitem{Sobolev} Sobolev S L 1962 {\it Soviet Math. Dokl.} {\bf 3}, 1307
\bibitem{Sobolev92} Sobolev S L 1992 {\it Cubature formulas and modern analysis} (Phyladelphia: Gordon and Breach Schience Publishers)
\bibitem{Ahrens} Ahrens C and Beylkin G 2009 {\it Proc. R. Soc.} {\bf 465} 3103
\bibitem{Haxton} Haxton D J 2007 {\it J. Phys.} B {\bf 40} 4443
\bibitem{Melezhik2016} Melezhik V S 2016 {\it EPJ Web of Conf.} {\bf 108} 01008
\bibitem{Burkardt} Burkardt J 2010 {\it Sphere Lebedev Rule, Quadrature Rules for the Unit Sphere}, Retreived from \url{https://people.sc.fsu.edu/~jburkardt/c_src/sphere_lebedev_rule/sphere_lebedev_rule.html}

\end{thebibliography}
\end{document}